\documentclass[twocolumn,showpacs,preprintnumbers,amsmath,amssymb]{revtex4}

\usepackage{epsfig,psfrag}
\usepackage{dcolumn}
\usepackage{bm}
\usepackage{graphicx}
\usepackage{color}

\begin{document}

\title{Nonequilibrium dephasing  in 
Coulomb blockade quantum dots} 

\author{Alexander Altland$^{1}$ and Reinhold Egger$^2$}
\affiliation{
${}^1$~Institut f\"ur Theoretische Physik, Universit\"at zu K\"oln, 
Z\"ulpicher Str. 77, D-50937 K\"oln, Germany\\
${}^2$~Institut f\"ur Theoretische Physik, 
Heinrich-Heine-Universit\"at, D-40225 D\"usseldorf, Germany}

\date{\today}

\begin{abstract}
We present a theory of  zero-bias anomalies and 
dephasing rates for a Coulomb-blockaded quantum dot, 
driven out of equilibrium by coupling to voltage biased source and
drain leads. We interpret our results in terms of the statistics of  voltage
fluctuations in the system.
\end{abstract}

\pacs{73.63.-b, 72.70.+m, 73.23.Hk}

\maketitle

Nonequilibrium phenomena encountered in steady-state quantum transport
through nanoscale or mesoscopic conductors are presently attracting a
lot of interest.  Recent experiments on mesoscopic wires have shown
that the nonequilibrium distribution function evolves from a two-step
to a broad single-step distribution as the effective interaction
strength increases \cite{saclay}. These experiments have led to rich
and nontrivial insights into dephasing and interaction physics in
general.  In the presence of a finite current flowing through the
system, noise will be generated, which then leads to decoherence and
dephasing processes. This phenomenon has been studied in some detail
for the nonequilibrium Kondo problem, both experimentally
\cite{kondo-exp} and theoretically \cite{rosch}, and a
zero-temperature decoherence rate $\Gamma(V)\sim V/\ln^2(V/T_K)$ was
reported in the regime $V\gg T_K$.  (Here $V>0$ is the applied bias
voltage, $T_K$ the Kondo temperature, and we set $\hbar=e=k_B=1$.)
Similar dephasing rates were found for a nonequilibrium Fermi-edge
singularity problem \cite{muzy}, and for a spin-fermion model driven
out of equilibrium \cite{mitra}.  Related questions are also
considered for disordered interacting quantum wires (Luttinger
liquids) \cite{luttinger}, and in the context of electronic
interferometry \cite{marquardt}.  Nonequilibrium dephasing rates can be
conveniently inferred from the voltage-induced broadening of the
zero-bias anomaly (ZBA) \cite{mirlin}, i.e. a smearing of the dips in
the energy-dependent tunneling density of states (TDoS), see also
Refs.~\cite{kamenev-andreev,rollbuehler}. (Note that in general,
dephasing rates will depend on the quantity under consideration
\cite{luttinger}.)

In this work, we present a theoretical study of the perhaps most
simple and basic scenario where one can study such questions, namely
for a metallic Coulomb-blockaded quantum dot \cite{aleiner}, with a
transport voltage $V$ applied via attached source and drain
electrodes. Note that the often-studied single-electron box (SEB)
\cite{schoen-zaikin,matveev,SEB,kamenev-gefen1} corresponds to a
single-lead situation plus a capacitively coupled gate, where one
cannot have steady-state current flow.  Previous theories for the
two-lead setup have only considered the $IV$ curve (or related
transport quantities) in certain limits \cite{zaikin,brouwer}.  As we
demonstrate below, the system is sufficiently simple to allow for a
quantitative study of nonequilibrium dephasing. At the
same time, it displays rich behavior that can be  probed
experimentally with available setups.

Nonequilibrium effects are captured by the Keldysh
formulation \cite{kamenev-andreev,kamenev}, where the total system
evolves from the initial time $-t_0/2$ after which the interaction is
smoothly switched on, to time $+t_0/2$ and back; one finally takes the
limit $t_0\to \infty$.  By a sequence of standard steps
\cite{kamenev-andreev,schoen-zaikin,lerner}, we obtain a  Keldysh functional
integral representation as 
\begin{equation}\label{genfunc}
\langle \hat X\rangle \equiv  \sum_{W\in \mathbb{Z}} \int {\cal D}(\phi,Q)
e^{iS_{\rm c}[\phi,Q]+ iS_{\rm tun}[\phi]} X[Q,\phi],
\end{equation}
where $X[Q_\sigma,\phi_\sigma]$ may be any observable expressed in
terms of the charge on the dot ($Q_\sigma(t)$) and its phase
($\phi_\sigma(t)$), both defined on the upper/lower ($\sigma=\pm$)
branch of the standard Keldysh contour~\cite{kamenev}. The phase
fields obey the boundary conditions, $\sum_{\sigma=\pm} \sigma
\phi_\sigma(-t_0/2)=2\pi W, \; \sum_{\sigma=\pm} \sigma
\phi_\sigma(+t_0/2)= 0$, where the integer $W$ is summed over
\cite{footwind}.  Here, $W$ is the real-time analogue of the winding
numbers central in  establishing Coulomb blockade
physics in imaginary-time theories \cite{SEB,kamenev-gefen1}.  The
charging energy, $E_c$, of the dot enters the theory through
\begin{equation}\label{charging}
S_{\rm c} = \sum_\sigma \sigma \int_{-t_0/2}^{t_0/2}
dt \left[ -E_c (Q_\sigma-Q_g)^2 + Q_\sigma \partial_t \phi_\sigma \right],
\end{equation}
where the constant $Q_g$ defines the electrostatically preferred
charge configuration. The coupling to the source and drain
($\alpha=\pm$) electrodes, biased by $\alpha V/2$, respectively, leads
to the tunnel action \cite{schoen-zaikin,zaikin}
\begin{equation}\label{stun}
S_{\rm tun} = \frac{ig_T}{2}\sum_{\sigma\sigma'}
\int dt dt' \ e^{-i\phi_\sigma(t)} \ L_{\sigma\sigma'}(t-t') \
e^{i\phi_{\sigma'}(t')} ,
\end{equation}
where, for simplicity, we assume identical tunneling conductances
$g_T$ for both contacts; the generalization to asymmetric cases is
straightforward.  The complex-valued functions $L_{\sigma\sigma'}(t)$
appearing in Eq.~\eqref{stun} are
\begin{equation}\label{ldef}
L_{\sigma\sigma'}(t)= \int_{-\infty}^\infty \frac{d\omega}{2\pi}
e^{-i\omega t} \left[ \sigma \sigma' 
L'(\omega) + (\sigma-\sigma') \omega \right],
\end{equation}
where the applied voltage $V$ and the temperature $T$ 
enter only via the real part,
\begin{eqnarray}\label{l1def}
L'(\omega)=  \omega \coth\left(\frac{\omega}{2T}\right)  
+ \sum_{\alpha=\pm}\frac{\omega+\alpha V}{2} \coth\left(\frac{\omega+\alpha
V}{2T}\right).
\end{eqnarray}
Our observable of main interest, the energy-dependent TDoS
$\nu(\epsilon,V,T)$ (technically, $(-\pi^{-1})$ times  the
imaginary part of the Keldysh retarded Green function) affords the
representation $\nu=\nu_e + \nu_h$, where $\nu_e$ ($\nu_h$) denotes the
contribution of electrons (holes) tunneling onto the dot,
\begin{equation} \label{tdos}
\frac{\nu_e(\epsilon)}{\nu_0} = {\rm Re}\int
d\tau\,e^{i\epsilon\tau} (1-n)(\tau) \left\langle 
e^{i(\phi_-(\bar t + \tau) -\phi_+(\bar t))} \right\rangle. 
\end{equation}
Here, $\nu_0$ is the noninteracting DoS, $\bar t$ is an arbitrary
reference time, $n(\tau)$ denotes the Fourier transform of the
double-step distribution function, $n(\epsilon) = \frac12
 (n_f(\epsilon+V/2) + n_f(\epsilon-V/2)),$ and
$n_f(\epsilon)=(e^{\epsilon/T}+1)^{-1}$ is the Fermi distribution
function.  The hole contribution obtains by exchange
$(1-n)\leftrightarrow n$ and $\phi_\pm \leftrightarrow \phi_\mp$. We
finally note that the current flowing through the dot is given by
\begin{equation}\label{iv}
I(V)=\frac{g_T}{2} \int d\epsilon [n_f(\epsilon-V/2)-n_f(\epsilon+V/2)]
\frac{\nu(\epsilon)}{\nu_0}, 
\end{equation}
with $I=I_0=(g_T/2) V$ in the absence of interactions ($E_c=0$).  The
phase factors in Eq.~\eqref{tdos} encapsulate the effects of particle
interactions.  We next explore the effect of these phase fluctuations
on the TDoS for the case $g_T\gg 1$, but return to the opposite limit,
$g_T\ll 1$, at the end of the paper.

\begin{figure}
\includegraphics[width=0.5\textwidth]{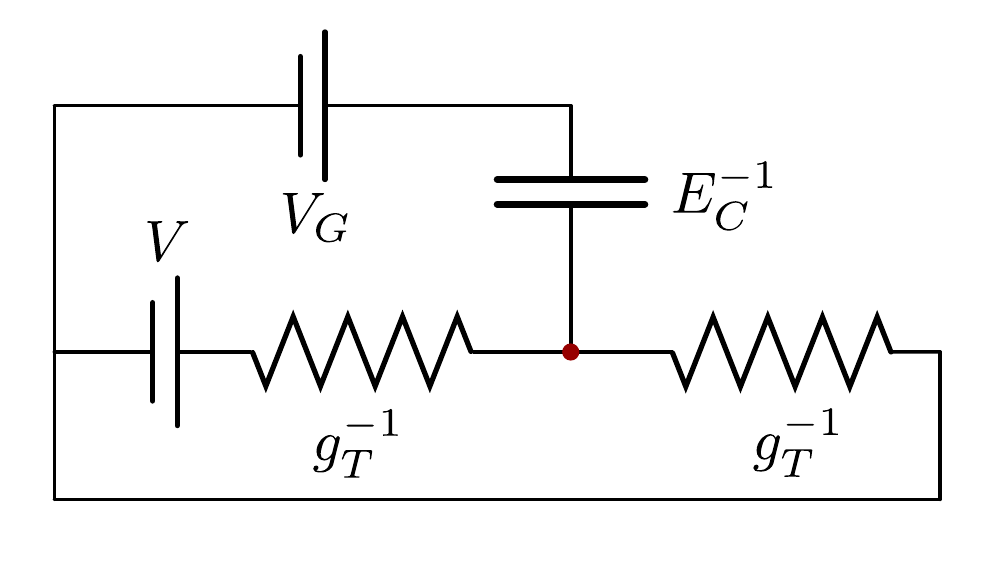}
\caption{\label{fig1} 
Classical resistor network equivalent of the system. 
For a  discussion, see main text. }
\end{figure}

In the ``open'' limit, $g_T\gg 1$, charge fluctuations are large, and
the phase representation obtained by integrating out the charge fields
in Eq.~\eqref{genfunc} can be restricted to a quadratic approximation
in the zero winding number sector, $W=0$.  Interaction effects are
then relatively weak, as in the corresponding SEB problem \cite{SEB},
and the dot distribution function remains close to the double step
$n(\epsilon)$. In this limit, the ``collective variables''
$\phi_\sigma$ essentially describe the classical non-equilibrium
steady state of a biased RC circuit, see Fig.~\ref{fig1}.  To explore
this connection, we expand the phase functional to second order in the
Keldysh ``classical'' and ``quantum'' fields, $\phi_c \equiv (\phi_+ +
\phi_-)/2$ and $\phi_q\equiv \phi_+-\phi_-$, respectively. A
``Hubbard-Stratonovich'' transformation of the contribution of ${\cal
 O}(\phi_q^2)$ then leads to the partition function $Z \simeq \int
{\cal D}(\phi_c,\phi_q) \left\langle e^{i\int dt
   \phi_q\left[(-C \partial_t^2 - 2g_T\partial_t)\phi_c +
     \xi\right]}\right\rangle_\xi,$ where we identified $E_c\equiv
(2C)^{-1}$ with an inverse classical capacitance. The
Hubbard-Stratonovich noise field $\xi$ with $\langle
\xi(t)\rangle_x=0$ is correlated as $\langle \xi(t) \xi(t') \rangle_x
= g_T L'(t-t')$, and $L'(t)$ is the Fourier transform of the kernel
(\ref{l1def}). Integration over $\phi_q$ then constrains fluctuations
of the voltage $U\equiv \partial_t \phi_c$ (relative to the stationary
value $V/2$) to solutions of the semiclassical $RC$-Langevin equation
$C\partial_t U(t) + R^{-1} U(t) =\xi(t)$, where $R=1/(2g_T)$ is the
parallel resistance of the circuit in
Fig.~\ref{fig1}, and   the fluctuating noise current
$I_{\Delta t} \equiv (2\Delta t)^{-1}\int_t^{t+\Delta t} dt'
\,\xi(t')$, averaged over a characteristic time window $\Delta t$,
gives rise to the dc noise power $S\equiv 2\Delta t \ \mathrm{var}(I)$
\cite{blanter}.  At finite temperatures and zero bias, the classical
limit of Eq.~\eqref{l1def}, $L'(t) = 4T \delta(t)$, describes a
thermal equilibrium situation, where the voltage $U(t)$ relaxes to the
Boltzmann distribution $P(U) \sim \exp(-CU^2/2T)$, and
\textit{Johnson-Nyquist thermal noise}, $S=4 T (g_T/2)$, is recovered.
For high voltages and low temperatures, $T\ll V$, the noise correlator
(\ref{l1def}) instead asymptotes to $L'(t) = V \delta(t)$, which
implies \textit{shot noise}, $S= 2 F I_0$, with the expected
\cite{blanter} Fano factor $F=1/2$:  upon
increasing $V$, transport through the quantum dot undergoes a
crossover from an equilibrium thermal to a steady-state nonequilibrium
shot-noise dominated regime.

Employing the TDoS (\ref{tdos}) as a reference observable, we next
explore the dephasing influence that this noise has on the quantum
physics of the system.  To this end, we use $\phi_c(t_2)-\phi_c(t_1) =
\int_{t_1}^{t_2} dt\, U(t)$ in Eq.~\eqref{tdos} and integrate over
$\phi_q$.  As a result, the (electron contribution to the) TDoS
assumes the form
\begin{align} \label{tdos_langevin}
\frac{\nu_e(\epsilon)}{\nu_0} = {\rm Re}\int
d\tau\,e^{i\epsilon\tau}(1-n)(\tau) \left\langle e^{i \int_{\bar
      t}^{\bar 
    t + \tau} dt U_e(t) }\right\rangle_\xi, 
\end{align}
where $U_e$ is the solution \cite{foot0} to a variant of the previous
Langevin equation,
\begin{equation} \label{langevin}
(C \partial_t + R^{-1})U_e=\xi + E_c
\left[ \delta(t-\bar t)+ \delta(t-(\bar t+\tau))\right]. 
\end{equation}
Equation \eqref{tdos_langevin} affords an intuitive interpretation:
the TDoS $\nu_e$ probes tunneling processes where an electron enters
the dot at time $\bar t$ and leaves at time $\bar t+\tau$.  In the
present structureless environment (a ``dot''), the dynamical phase
controlling these processes obtains by integration of the
time-dependent voltage on the dot. The latter is governed by a
superposition of (i) a voltage pulse of height $E_c$ upon the entry of
the external particle, as described by the second term on the
r.h.s. of Eq.~\eqref{langevin} \cite{foot1}, and (ii) the background
voltage noise $\xi$ on the dot. Averaging the exponentiated solution
of (\ref{langevin}) over the $\xi$ and adding the hole contribution,
$\nu_h$, we obtain
\begin{eqnarray}\label{large}
\frac{\nu(\epsilon)}{\nu_0} &=& 1 - \frac{1}{4\pi g_T} 
\int_0^\infty
d\tau\sum_{\alpha=\pm}\frac{\cos[(\epsilon+\alpha
 V/2)\tau]}{\tau}  \\ \nonumber
&\times& \left(1-e^{-\Omega\tau} \right) e^{-S(\tau)} 
,
\end{eqnarray}
where non-singular contributions of higher order in $g_T^{-1}$ have
been neglected. In Eq.~(\ref{large}), the first term under the integral is
the temporal Fourier transform of the distribution function, the
factor $(1-e^{-\Omega t})$, $\Omega\equiv (RC)^{-1}$ accounts for the
relaxation of the initial voltage pulse, and the last term defines the
average noise-action (cf.~Eq.~\eqref{l1def}),
\begin{equation}\label{staudef}
S(\tau) = \frac{\Omega^2}{4\pi g_T}\int_{0}^\infty
d\omega \frac{1-\cos(\omega\tau)}{\omega^2(\omega^2+\Omega^2)} L'(\omega).
\end{equation}
Note that in this weak Coulomb blockade limit, the TDoS is independent
of the reference charge $Q_g$.  In what follows, we consider the limit
$T=0$ where the influence of nonequilibrium dephasing on the TDoS is
strongest.

\begin{figure}
\includegraphics[width=0.45\textwidth]{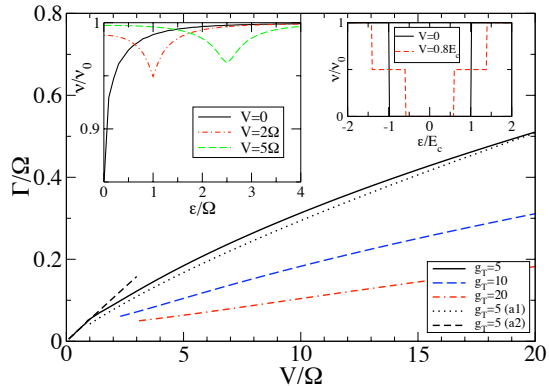}
\caption{\label{fig2} (Color online) Dephasing rate vs voltage (at
 $T=0$) in units of the inverse RC time, $\Omega=4g_T E_c$, for
 different values of $g_T$. Curves were obtained by numerical
 integration of Eq.~\eqref{decohlarge}.  Dotted (a1) and dashed (a2)
 curves represent the analytical predictions in Eqs.~\eqref{ana1} and
 \eqref{ana2}, respectively; for (a2), $c=0.25$ in Eq.~\eqref{ana2}.
 The insets show the TDoS $\nu(\epsilon)/\nu_0$ at $T=0$ as a
 function of $\epsilon$ for several $V$. Left inset: from numerical
 integration of Eq.~\eqref{large} for $g_T=5$.  Right inset: from
 Eq.~\eqref{tdo2} for $g_T\ll 1$ and $Q_g=0$.}
\end{figure}

To logarithmic accuracy, the equilibrium ($V=0$) limit of
Eq.~\eqref{staudef} can be approximated by $S_0(\tau)\simeq (2\pi
g_T)^{-1}\ln(\Omega \tau)$. The resulting TDoS is symmetric in
$\epsilon$ and displays a characteristic dip at $\epsilon=0$ -- the
ZBA -- which vanishes in the limit $E_c\to 0$, and also for $g_T\to
\infty$. At finite voltage, the TDoS remains symmetric, while the
double-step profile of the distribution $n(\epsilon)$ implies a
splitting of the $\epsilon=0$ ZBA into two minima at $\epsilon=\pm
V/2$, see Fig.~\ref{fig2}.  The strict positivity of
$S(\tau)-S_0(\tau)>0$, previously identified as a manifestation of
enhanced noise levels for $V\not=0$, then causes a suppression of the
ZBA. At the same time, its line shape broadens. In the limit $V\to
\infty$, the linear growth $S(\tau)\sim V\tau$ implies a vanishing
ZBA.  Similar phenomena were recently studied~\cite{mirlin} for
tunneling into a 2D diffusive metal. Eq.~(\ref{large}) may serve to
define a \textit{nonequilibrium dephasing rate} $\Gamma(V)$ from the
voltage-induced broadening of the ZBA dips at $\epsilon=\pm
V/2$~\cite{mirlin}.  Writing $\delta\epsilon=\epsilon-V/2$ with
$|\delta\epsilon|\ll \Omega$, the ZBA dip is described by $\delta
\nu(\delta \epsilon) =\nu_0 - \nu(\epsilon)$.  Parameterizing small
deviations off the dip at $V/2$ in terms of the dephasing rate
\cite{foot2}, $ [\delta\nu(0)-\delta\nu(\delta\epsilon)]/\delta\nu(0)
\simeq \frac12 [\delta \epsilon/\Gamma(V)]^2,$ Eq.~\eqref{large}
yields
\begin{equation}\label{decohlarge}
\Gamma^{2}(V) =  \frac{\int_0^\infty \frac{d\tau}{ \tau }
e^{-S(\tau)} (1-e^{-\Omega\tau})}{\int_0^\infty d\tau \ \tau
e^{-S(\tau)} (1-e^{-\Omega\tau})} .
\end{equation}
Analytical results for $\Gamma(V)$ can then
be extracted from Eq.~\eqref{decohlarge} in various limiting regimes,
while the full curve is obtained numerically, see Fig.~\ref{fig2}.

Let us first discuss the asymptotic regime of weak interactions,
$E_c\to 0$. (Technically, this means that we take the limit $\Omega\to
0$ prior to $\delta \epsilon\to 0$.) Eq.~\eqref{decohlarge} is then
dominated by contributions $\Omega \tau\ll 1$, which implies
$S(\tau)\approx V E_c \tau^2/4$ and $\Gamma(V)= \sqrt{E_c V/2}$. The
$g_T$-independence of $\Gamma$ reflects the diverging $RC$-time:
particles tunneling onto the dot do have time to realize the
dissipative nature of their environment. Turning to the complementary
regime, $\delta\epsilon\to 0$ at fixed $\Omega$, we note that the leading
contribution to an expansion of the TDoS in $g_T^{-1}\ll 1$, obtained
by setting $S=0$, is given by the logarithmically singular expression
$\delta\nu(\delta\epsilon)/\nu_0 = (8\pi g_T)^{-1} \ln
[1+(\Omega/\delta\epsilon)^2]$.  Using Eq.~\eqref{iv}, this recovers
the well-known result \cite{zaikin,brouwer} for the differential
conductance, $(2/g_T) dI/dV = 1 - (4\pi g_T)^{-1}
\ln[1+(\Omega/V)^2].$ However, in order to define the dephasing rate,
one needs to retain a finite noise action $S$.  Specifically, for
$V>\Omega$, $S(\tau)\simeq V\tau/(4\pi g_T)$, implying
\begin{equation}\label{ana1}
\frac{\Gamma(V)}{V} \simeq \frac{1}{4\pi g_T} \left[
\frac{\ln( 1+4\pi g_T \Omega/V )}{ 1 - (1+4\pi g_T\Omega/V)^{-2} } 
\right]^{1/2},
\end{equation}
while for $\Omega e^{-\pi g_T}<V<\Omega$ \cite{foot3}, 
we estimate
\begin{equation}\label{ana2}
\frac{\Gamma(V)}{V} \simeq \frac{c}{\sqrt{2\pi g_T}}
(V/\Omega)^{-1/(4\pi g_T)},
\end{equation}
where $c$ is a numerical prefactor of order unity.  Importantly, in
all parameter regimes, we find $\Gamma(V)<V$, and the double peak
structure in the nonequilibrium TDoS is reasonably well resolved.  As
a function of $V$, the relative strength of the dephasing rate
(i.e. the ratio $\Gamma/V$) increases with decreasing $V$.
Furthermore, Eq.~\eqref{ana1} shows that for very large voltage,
$V>g_T \Omega$, the rate $\Gamma(V)\sim V/g_T$ is directly determined
by the shot noise $S=2F I_0$ discussed above.  The analytical results
\eqref{ana1} and \eqref{ana2} describe the numerical solution to
Eq.~\eqref{decohlarge} rather well, as indicated in Fig.~\ref{fig2}.
With increasing $g_T$, the dephasing rates are gradually suppressed,
and for $V=0$, as expected, the dephasing rate is zero.

Finally, we briefly turn to the
opposite case of strong Coulomb blockade, $g_T\ll 1$, where
the dual charge representation is appropriate.  Expanding
$e^{iS_{\rm tun}}$ in Eq.~\eqref{genfunc} into a Taylor series, we can
integrate out the phase fields $\phi_\sigma$.  Now the $e^{\pm i
\phi_\sigma(t)}$ factors appearing in the tunnel action \eqref{stun}
can be interpreted as charge raising or lowering operators, i.e.  they
change the corresponding $Q_\sigma(t)$ variable by $\pm 1$.  In order
$m$ of the tunnel expansion, we thus have $2m$ jumps at times $t_j,
t'_j$ ($j=1,\ldots,m$), with jump directions $\sigma_j,\sigma'_j=\pm 1$.  
The charge fields are then expressed in terms of these jump
times and directions, 
$Q_\sigma(t)= N - \sigma \sum_{k=1}^m \left [\delta_{\sigma\sigma_k} 
\Theta(t-t_k)-\delta_{\sigma\sigma_k'}\Theta(t-t_k') \right],$
where $\Theta(t)$ is the Heaviside function and 
the $W$-summation in Eq.~\eqref{genfunc} implies the boundary condition
$Q_\pm(-t_0/2)=N\in \mathbb{Z}$. 
The charge representation of Eq.~\eqref{genfunc} is then given by \cite{foot4}
\begin{eqnarray}\label{charge}
Z &=& \sum_{m=0}^\infty \frac{1}{m!} \left(\frac{-g_T}{2}\right)^m
\sum_{ N,\{\sigma,\sigma'\} } \int \prod_{j=1}^m dt_j dt'_j 
\\ &\times& \nonumber
e^{-i\sum_\sigma \sigma E_c \int dt [Q_\sigma(t)-Q_g]^2} 
\prod_{k=1}^m L_{\sigma_k,\sigma_k'}(t_k-t_k').
\end{eqnarray}
Away from charge degeneracies, i.e.~for $Q_g$ not half-integer,
Eq.~\eqref{charge} admits a solution in terms of a
``noninteracting-blip approximation'' (NIBA) \cite{weiss}, where $Z$
is dominated by short ``blips'' of length $|t_j-t_j'|$ during which
$Q_+(t)\ne Q_-(t)$, with long time intervals between subsequent blips
where $Q_+(t)=Q_-(t)$.  Defining the optimal charge state $\tilde Q\in
\mathbb{Z}$ according to $Q_g-1/2<\tilde Q <Q_g+1/2$, the NIBA
transition rate from $Q_\pm=\tilde Q$ to any other charge state
\textit{vanishes} for $T=0$ and $V<2E_c|\tilde Q-Q_g+1/2|$.  In this
regime, the TDoS is thus given by (see Fig.~\ref{fig2})

\begin{equation}
\label{tdo2}
\frac{\nu(\epsilon)}{\nu_0} = 1 -\sum_{s=\pm,\alpha=\pm} \frac{s}{2}
\Theta\left(\epsilon+[s+2(\tilde Q-Q_g)]E_c+\frac{\alpha V}{2} \right),
\end{equation}
which recovers the equilibrium result of Ref.~\cite{kamenev-gefen1}
for $V=0$. Note that $\nu(\epsilon)$ is generally asymmetric;
Eq.~\eqref{iv} implies well-known Coulomb blockade expressions for the
$IV$ curve.  Vanishingly small (noise) current levels imply that
nonequilibrium dephasing is strongly suppressed in the Coulomb
blockade regime and may leave traces only beyond NIBA.  The
near-degeneracy case, on the other hand, is related to a many-channel
variant of the Kondo effect \cite{matveev}, where frequent transitions
between two charge states occur. NIBA does not hold in this regime,
but a real-time renormalization-group approach \cite{mitra} can be
constructed and will be described elsewhere.

This work was supported by the SFB Transregio 12 of the DFG, and by
the ESF program INSTANS.


\begin{thebibliography}{99}
\bibitem{foot2} 
In principle, there may also be a
linear term (asymmetric dip), which is of no interest in this context,
or even nonanalytic behavior at $V=0$.
\bibitem{foot3}
A quadratic expansion of the action in $\phi_c$,
and hence Eq.~\eqref{large}, breaks down for $V<\Omega e^{-\pi g_T}$, 
see Ref.~\cite{zaikin}.
\bibitem{foot4}
The normalization $Z=1$ is not obvious in Eq.~\eqref{charge}, but 
after $N$-summation, a cancellation among the different Keldysh 
sectors indeed implies $Z=1$.
\end{thebibliography}
\end{document}